\documentclass[num-refs]{wiley-article}

\usepackage{graphicx}
\usepackage{caption}
\usepackage{subcaption}

\usepackage{color}
\usepackage{ulem}

\newcommand{\delete}[1]{}

\newcommand{\mnote}[1]{}

\usepackage{siunitx}

\papertype{Original Article}
\paperfield{Journal Section}

\title{Fast, high precision autofocus on a motorised microscope: automating blood sample imaging on the OpenFlexure Microscope} 

\author[1]{Joe Knapper}
\author[1]{Joel T. Collins}
\author[1]{Julian Stirling}
\author[2]{Samuel McDermott}
\author[1]{William Wadsworth}
\author[1]{Richard Bowman}

\affil[1]{Centre for Photonics and Photonic Materials, Department of Physics, University of Bath, UK.}
\affil[2]{Cavendish Laboratory, University of Cambridge, UK.}

\corraddress{Richard Bowman, Centre for Photonics and Photonic Materials, Department of Physics, University of Bath, UK.}
\corremail{rwb34@bath.ac.uk}

\fundinginfo{EPSRC, Grant/Award Number: (EP/R013969/1 and EP/R011443/1; The Royal Society, Grant/Award Number: URF\textbackslash R1\textbackslash 180153}

\runningauthor{Knapper et al.}

\begin{document}

\maketitle

\begin{abstract}
The OpenFlexure Microscope is a 3D printed, low-cost microscope capable of automated image acquisition through the use of a motorised translation stage and a Raspberry Pi imaging system. This automation has applications in research and healthcare, including in supporting the diagnosis of malaria in low resource settings. The \textit{plasmodium} parasites which cause malaria require high magnification imaging, which has a shallow depth of field, necessitating the development of an accurate and precise autofocus procedure. We present methods of identifying the focal plane of the microscope, and procedures for reliably acquiring a stack of focused images on a system affected by backlash and drift. We also present and assess a method to verify the success of autofocus during the scan. The speed, reliability and precision of each method is evaluated, and the limitations discussed in terms of the end users' requirements.


\end{abstract}

\section{Introduction}

\begin{figure*}
   	\centering
	\includegraphics[width=0.6\textwidth]{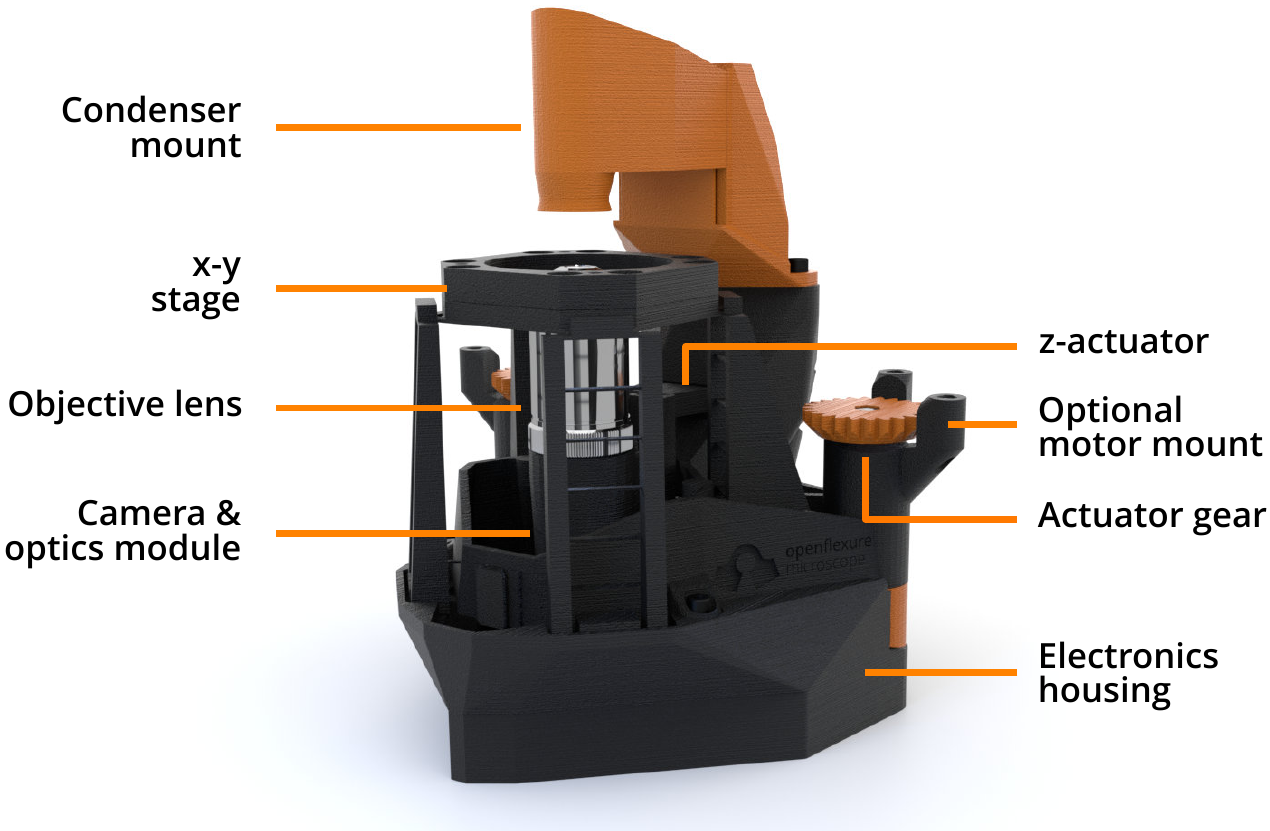}
	\caption{A labelled diagram of the OpenFlexure Microscope in a trans-illumination set up. Reproduced under CC-BY licence from Collins \textit{et al.} \cite{Collins2020}.}
	\label{figure:OFM}
\end{figure*}

As the number and accessibility of low-cost, locally manufactured microscopes continues to grow, their potential applications are also increasing. Microscopes costing less than 1 USD can be assembled and used in outreach and education \cite{Cybulski2014}, while more expensive designs have applications in research. An example of an open-source, lab grade microscope is the OpenFlexure Microscope, a 3D printed microscope capable of automated sample positioning and high resolution imaging \cite{Collins2020}. Costing 200 USD in parts, a research grade version of the OpenFlexure Microscope (shown in Figure \ref{figure:OFM}) combines a Raspberry Pi, Pi Camera and a custom motor controller board with a 3D-printed precision translation mechanism to precisely position and image a sample in a 12 $\times$ 12 $\times$ 4 mm volume. Flexure hinges are used to translate the stage and sample in $xy$, and the optics module in $z$. Standard RMS objectives can be screwed in to provide magnification, requiring an achromatic tube lens with a focal length of 50 mm.  Based on the user's requirements, the OpenFlexure software can be used to automate the capturing of large areas of a sample \cite{Collins2021}. This has the potential to improve the efficiency of research, as users can automate the sample movement and image collection conventionally performed by hand.

As with manual microscopy, automatically moving across a sample capturing images requires frequent refocusing, as hardware and sample imperfections move the sample away from the object plane. Conventional microscopes require technicians to be trained in how to identify and return to the desired focal point, adjusting the height of the sample or optics until the image becomes as sharp as possible. Automated microscopy requires a similar procedure, moving the sample while tracking position and image sharpness before returning to the position of highest sharpness. As the depth of field of an objective decreases as magnification increases, higher power microscopes require more frequent and more precise autofocus.

One application of microscopy requiring high magnification is in healthcare, where blood, cerebrospinal fluid and other samples are examined for evidence of parasites or bacteria \cite{Bennett2009}. The size of disease-causing parasites are often within an order of magnitude of the diffraction limit (\SI{0.2}{\micro\meter}) of high NA objectives. This necessitates the highest levels of magnification possible, typically employing a 1.2 NA, 100$\times$ oil immersion objective. These objectives have a depth of field below \SI{1}{\micro\meter}, and so require extremely accurate focus. An autofocus not sufficiently accurate or reliable would produce unusable out-of-focus images, increasing the chance of a misdiagnosis.

High-power automated microscopes require a rapid autofocus procedure suitable for a narrow depth of field. On the OpenFlexure Microscope, the procedure must be designed to accommodate limited computational resources, backlash and errors in positioning. In this paper, a modified 2D Laplacian and JPEG file size are assessed as image sharpness metrics, based on their reliability, speed and suitability to the samples of interest. The optimal order of movements and recalibration measurements are discussed in terms of a closed-loop, self-correcting solution to imperfect positioning hardware. These measurements and movements are combined into several autofocus options, allowing the user to make an informed choice based on their required speed, reliability and sample type.

\section{Background and Motivation}

The recent increase in affordable hardware and accessible manufacturing methods has lead to a surge of interest in low-cost microscopy. Microscopes suitable for a range of budgets and applications have been developed and shared as open-source projects, including Foldscope, Microscopi and UC2 \cite{Cybulski2014, Wincott, Diederich2019}. The primary applications of low-cost microscopes are in education and outreach, allowing users to quickly and easily view samples.

In sub-Saharan Africa, where malaria is endemic, manual light microscopy is considered to be the ``gold standard'' of diagnosis \cite{WHO2000}. Microscopy performed on a blood sample is sensitive, rapid and can also diagnose other conditions. Although the cost per test scales favourably, the cost of supplying low resource areas with equipment, training and maintenance is considerable, motivating research into a more accessible alternative. To declare a thick smear blood sample as negative for malaria, WHO training requires 100--200 fields of view to be examined \cite{WHO2010}. Low-cost microscopes with automated sample movement and image capturing present a viable alternative to the time-consuming process of manually searching a sample. Low-cost microscopes enabled by local manufacturing and supported by a community of users have several advantages over commercial microscopes.

Manufacturing the microscope locally greatly decreases the upfront and maintenance costs, and is less reliant on international supply chains. Low-cost microscopes are generally lighter and more portable than their commercial counterparts, increasing the areas which can offer the testing. This makes rapid, point-of-care (POC) diagnosis more accessible, reducing delays in treatment or the opportunities for administrative errors.

Malaria diagnosis requires high magnification imaging, usually employing a 100$\times$ oil immersion objective with an NA of at least 1.2. These objectives have a depth of field less than \SI{1}{\micro\meter}, requiring the sample, optics and camera to be precisely positioned to focus the image. For a low-cost, automated microscope to be used in the image acquisition for malaria diagnosis, the autofocus procedure needs to reliably position the sample in the small range in which it is sufficiently focused. Even slightly defocused images, or areas of the sample fully out-of-focus, limit the diagnostician's access to information. The consequences of a misdiagnosis in these cases can be severe. In addition to the health risks of a patient receiving incorrect treatment, incorrect diagnoses have been shown to have wider-reaching societal and economic impacts. Public trust in healthcare can be reduced, and valuable medication can be wasted treating the wrong condition.

OpenFlexure software including a simple autofocus procedure was trialled by our collaborators at the Ifakara Health Institute (IHI). Using blood samples already collected and tested by their standard procedure, images were collected as follows. At each xy position in a user-defined grid, an autofocus procedure (details given in Section \ref{section:moving}) was performed to position the objective such that the sample was at the focal point. The objective was then moved a fixed distance below the focal point, and an image captured. The objective was then raised in regular steps (d$z$), capturing an image at each point. This series of images is known as a z-stack, and should be centred on the focal plane. The centre image is used for diagnosis, while the images taken above and below are used later to assess how well the system focused.

\begin{figure}[t]
   	\centering
   	\includegraphics[width=0.45\textwidth]{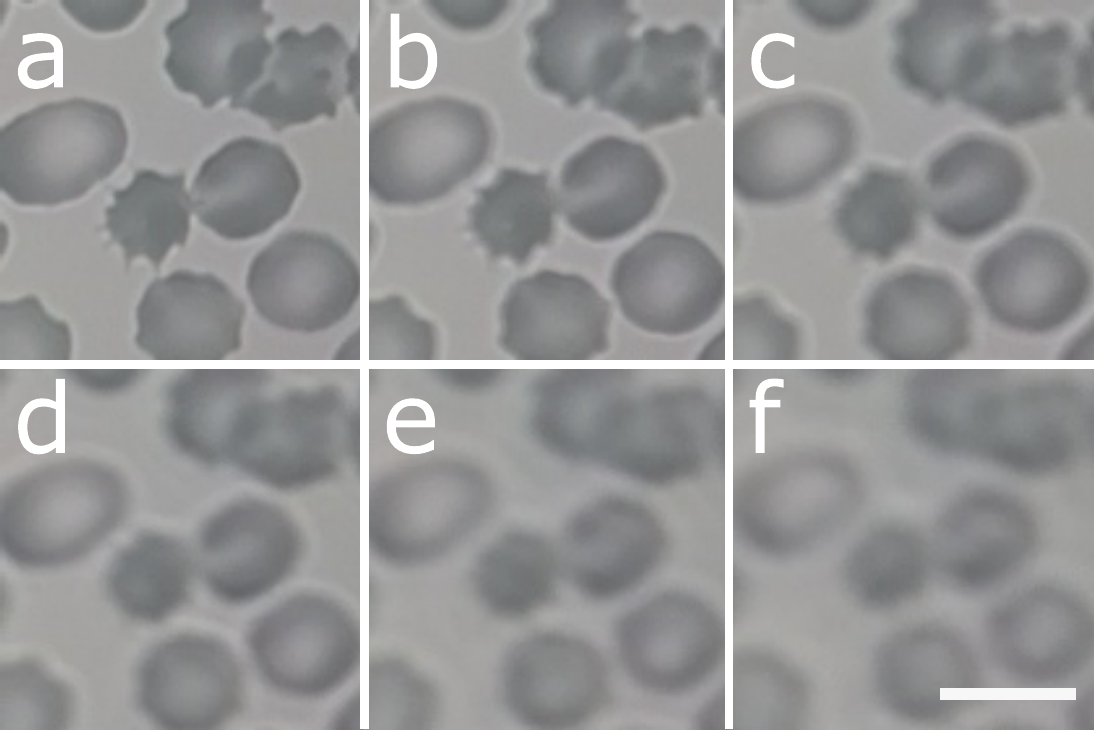}
	\caption{Cropped blood sample images taken on the OpenFlexure Microscope using a 100$\times$ oil immersion objective with an NA of 1.2, at distances \SI{0}{\micro\meter} \textbf{(a)}, \SI{1}{\micro\meter} \textbf{(b)}, \SI{2}{\micro\meter} \textbf{(c)}, \SI{3}{\micro\meter} \textbf{(d)}, \SI{5}{\micro\meter} \textbf{(e)} and \SI{10}{\micro\meter} \textbf{(f)} from the focal point. Scale bar is \SI{10}{\micro\meter}.}
	\label{figure:scales}
\end{figure}

Stacks of 5, 7 and 9 images were taken, with image spacing ranging from \SI{0.5}{\micro\meter} to \SI{5.0}{\micro\meter}. Assessing these z-stacks showed that the accuracy in positioning required for high magnification focus necessitated closed loop correction steps. In particular, the changing of direction at the beginning of the z-stack introduced more uncertainty in position than the precision required for diagnosis. Figure \ref{figure:scales} shows the level of blur introduced by positioning the sample 1, 2, 3, 5 and \SI{10}{\micro\meter} away from the focal point.

\section{Measuring sharpness}

Autofocus procedures are commonly used in digital cameras and smartphones, adjusting the focus to produce the sharpest image. Sharpness metrics are also commonly used to assess the effect of modifications to imaging systems, such as in adaptive optics \cite{Gould2013}. There are numerous sharpness metrics available for digital images, generally giving higher scores to images with rapidly changing colours, hard lines and high frequency information. Sun \textit{et al.} proposed five criteria for assessing the performance of a sharpness metric in digital microscopy \cite{Sun2004}. The criteria, when testing multiple sharpness metrics on the same stack of images, can be used to predict the performance of the metric on similar samples. These criteria include the proximity of the metric's sharpness peak to the known sharpest image, and the width of the peak. The further criteria assess the number, size and distance to secondary peaks. For many applications, speed and computational resources are less important. When performed on a Raspberry Pi however, the speed and how computationally expensive the metric is are also important factors.

When focusing the OpenFlexure Microscope, sharpness metrics are used to give a relative measure of image sharpness. These measures also award a higher score to images with more features, and so an individual image score isn't sufficient to judge if an image is sparse and well-focused, or densely populated and out of focus. Due to the lack of a reference image for comparison, in image quality assessment, this is a no-reference problem. Metrics can only assess based on the data they collect, rather than having prior knowledge of how a focused image should look. In addition to assessing the performance of a sharpness metric, its suitability for various samples and likely causes of failure must be understood. As the OpenFlexure Microscope has the option for down-sampled (832 $\times$ 624 pixels, $\sim$300 KB) or full resolution (3280 $\times$ 2464 pixels, $\sim$5 MB) captures, sharpness metric performance was assessed for each.

\begin{figure}[t]
   	\centering
	\includegraphics[width=0.46\textwidth]{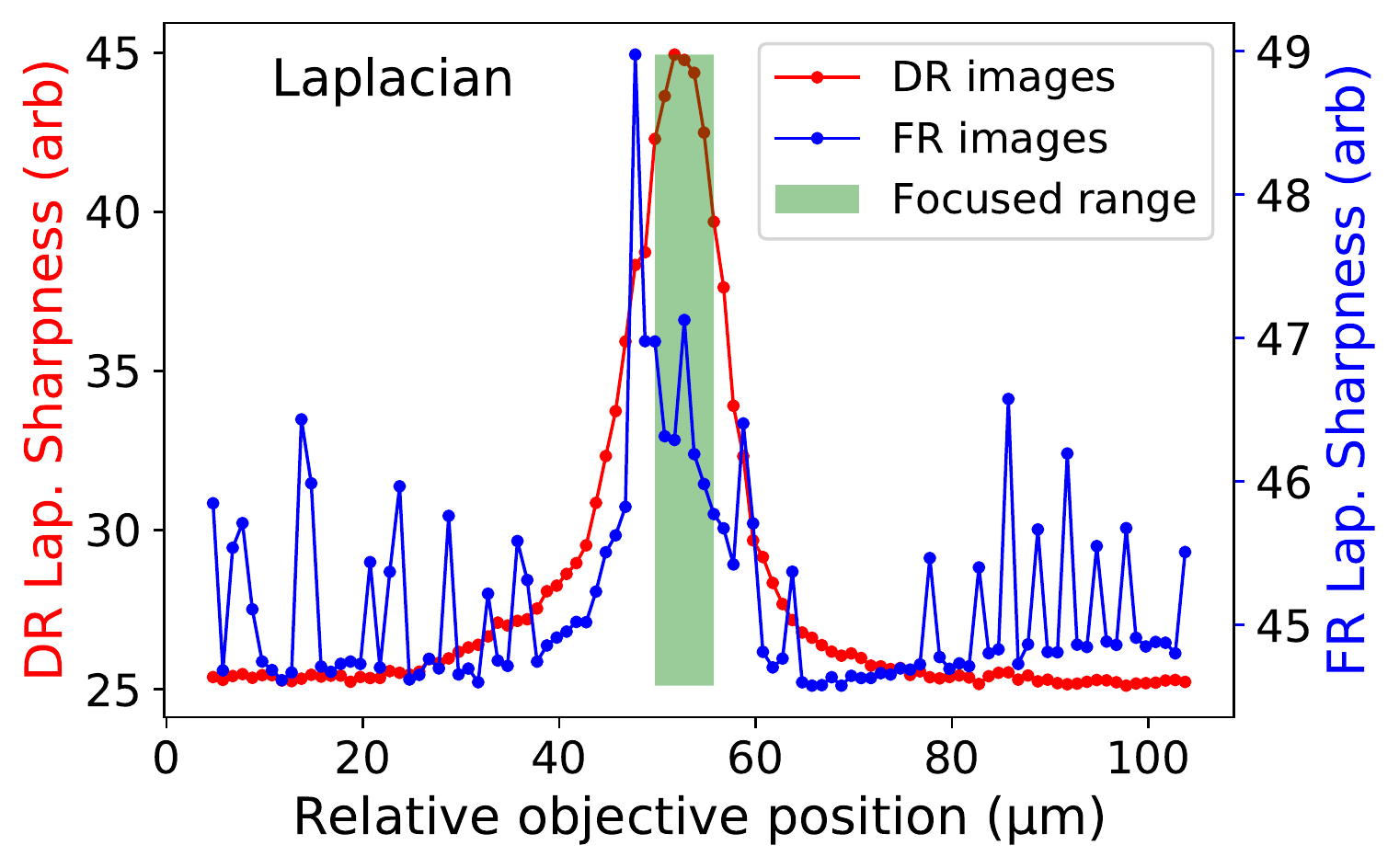}
	\includegraphics[width=0.49\textwidth]{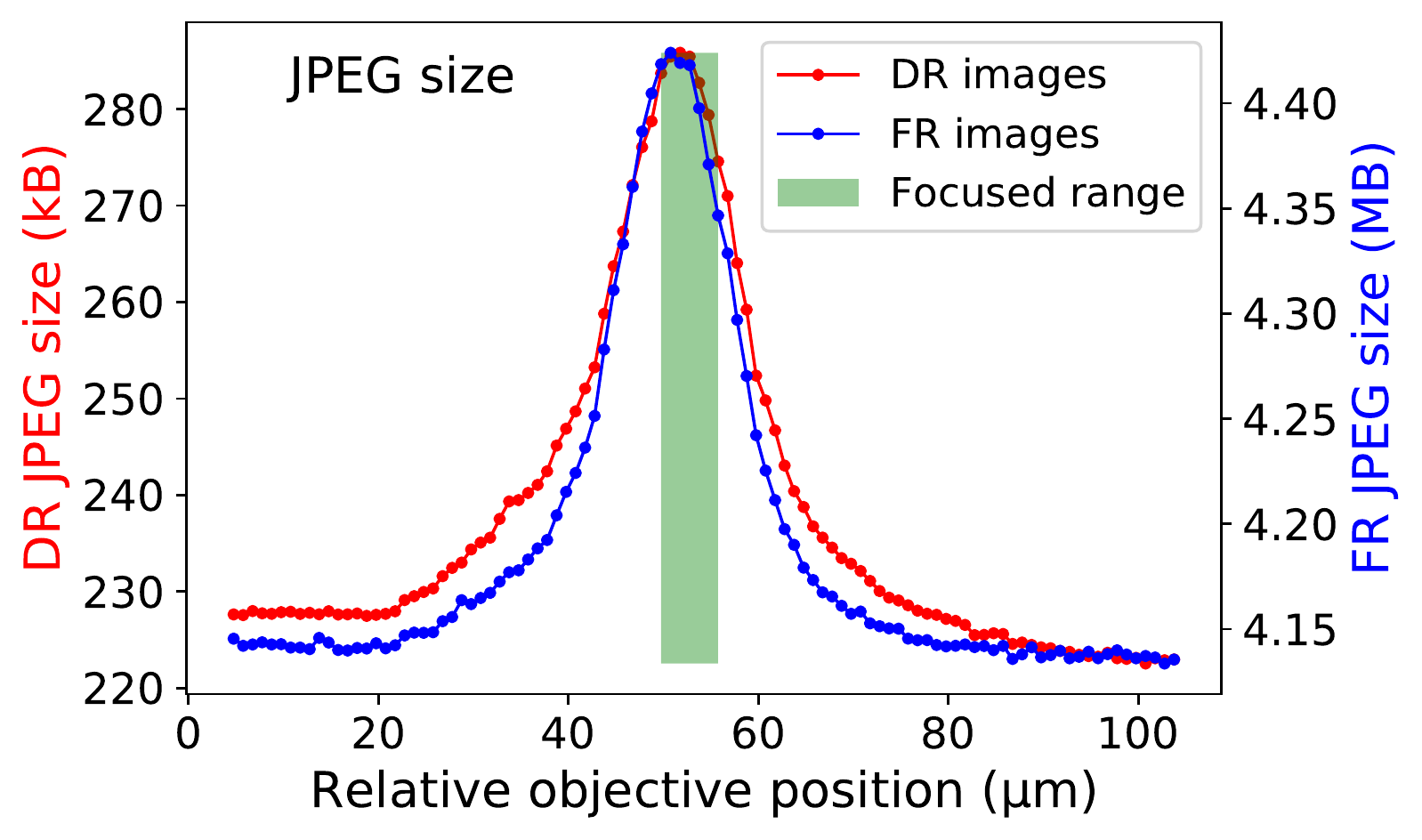}
	\caption{Plots of sharpness measures against height for the same stack of 99 down-sampled resolution (DR) and full resolution (FR) images of a \textit{zea} seed sample. The green zone represents a manually assessed range of in-focus images. \textbf{(a)} uses the modified Laplacian metric, and \textbf{(b)} uses the JPEG file size. Each step is \SI[separate-uncertainty = true, multi-part-units = single]{50(2)}{\nano\meter}.}
	\label{figure:sharpness}
\end{figure}

A modified 2D Laplacian is commonly used for edge detection and to assess image sharpness \cite{Hao2015}. The form used in OpenFlexure software converts a captured image to greyscale before convolving the image with a Laplacian kernel \cite{Virtanen2020}

\begin{equation}
    \begin{bmatrix}
        0 & 1 & 0\\
        1 & -4 & 1\\
        0 & 1 & 0
    \end{bmatrix}.
\end{equation}

\noindent This score of colour change around the pixel is squared and averaged over the image to give an image sharpness score. The image in a stack with the highest Laplacian score was taken to be the most focused, as hard edges and sharp features are characteristic of a focused image.

An autofocus procedure using this method was written and tested on an ipomoea root sample. The microscope captured a z-stack of 7 images equally spaced throughout the range to be searched. The Laplacian sharpness of each image was assessed, and the objective returned to the position of greatest sharpness. A trial repeating this autofocus procedure ten times on an ipomoea root sample found that focusing took an average of 14.9 seconds with a standard deviation of 1.0 s. As a typical scan on blood samples requires 100 $x$-$y$ positions be focused on, at this rate, refocusing would add over 24 minutes to each scan, reducing throughput and delaying results.

An alternative sharpness metric which reduces this delay utilises the OpenFlexure software's MJPEG preview stream, included to aid the user in positioning samples and previewing captures. With bit-rate control disabled (as in OpenFlexure software), an MJPEG stream consists of independent JPEG frames. JPEG encoding splits the image into 8 $\times$ 8 pixel blocks, describing each block using superpositions of discrete cosine functions \cite{Wallace1992}. To save storage space, each block is described using the fewest possible cosine functions. Focused images with high frequency information and hard boundaries will require more cosine functions to describe them than an out-of-focus, blurred image. This causes sharper images to have a larger file size, increasing the data rate required to maintain the MJPEG stream. By tracking this size and the $z$ position as the objective is swept through the focal point, a peak sharpness can be identified and returned to.

A fast autofocus procedure based on this metric was written and included in OpenFlexure software \cite{Collins2021}. The sample moves through a range of interest (typically \SI{100}{\micro\meter}) and the JPEG size is tracked. Unlike the Laplacian metric, this method is computationally `free', as the MJPEG stream is generated in the GPU by default. This enables sample positioning to be assessed in real time, allowing the sample to continue moving throughout the measurements. This allows more positions to be assessed, with the spacing between images limited only by the frame rate and movement speed. Ten trials of this procedure on the same ipomoea root sample took an average of 4.1 s with a standard deviation of 0.4 s. Across a 100-location blood scan, this reduces the time taken focusing from 24 minutes to under 7.

\begin{figure*}[t]
   	\centering
   	\includegraphics[width=0.45\textwidth]{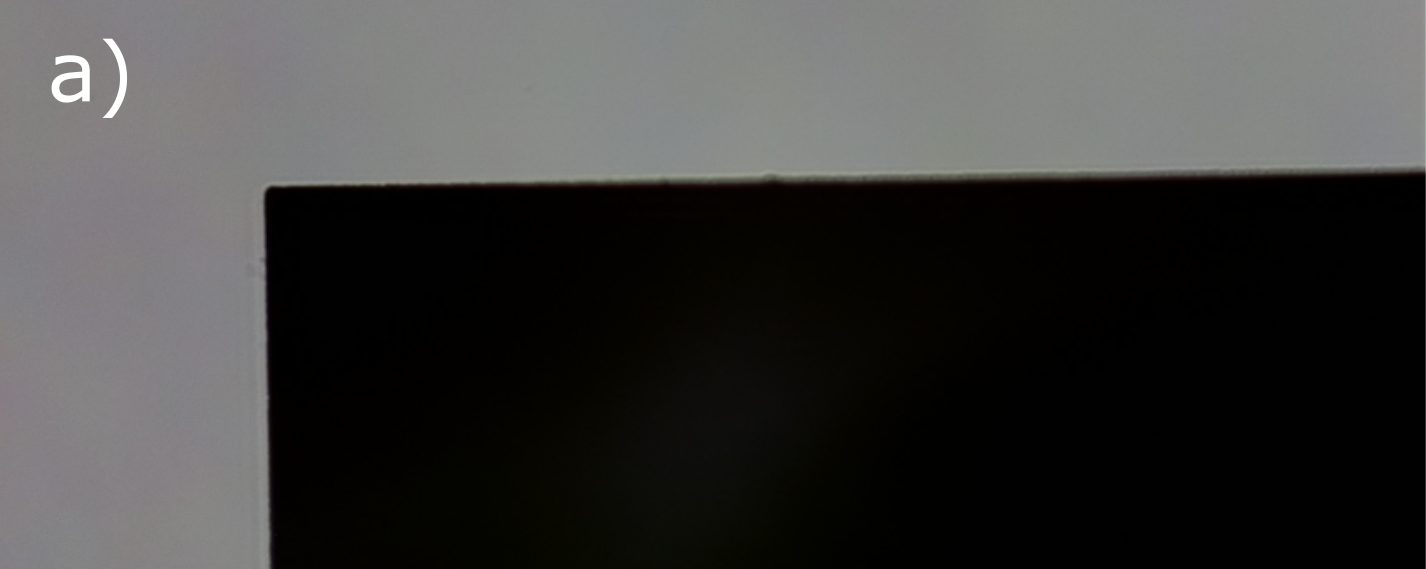}
   	\includegraphics[width=0.45\textwidth]{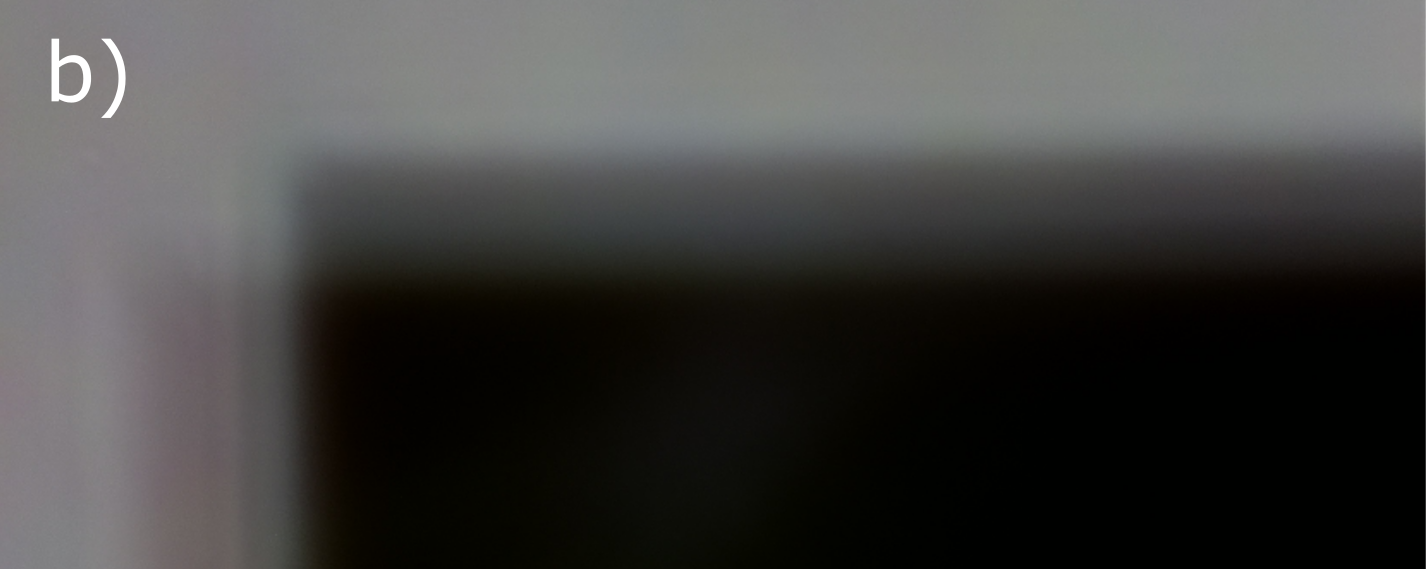}
    	\caption{Images of a 1951 USAF resolution target edge taken on the OpenFlexure Microscope using a 100$\times$ oil immersion objective with an NA of 1.2. \textbf{a)} is taken at the focal point, and \textbf{b)} taken {\SI{51}{\micro\meter}} away. The defocusing causes a more gradual edge, but introduces information into 8$\times$8 pixel blocks that would otherwise be empty, increasing the JPEG size.}
    	\label{figure:USAF}
\end{figure*}

The performance of these metrics were assessed by post-processing a stack of 99 images of a zea seed section, taken $\sim$ \SI{1}{\micro\meter} apart. Images were captured at both down-sampled and full resolution on a Raspberry Pi camera, with magnification provided by a 40$\times$, 0.65 NA RMS objective. The focused range was judged manually from examination of the images in the stack, and covers a range of approximately \SI{5}{\micro\meter}. We note this is thicker than the $\sim$ \SI{1}{\micro\meter} depth of field (DOF) of the objective used. This is due to the thickness of the sample and subjectivity in determining when the images are considered focused.

The Laplacian and JPEG sharpness metrics were applied to each image in the stack, and the results are shown in Figure \ref{figure:sharpness}. The position of the highest peak for the full resolution Laplacian in Figure \ref{figure:sharpness} motivated an examination of the image corresponding to that position. This examination revealed an area of approximately 590 $\times$ 1 pixel in which the capture failed, causing a black line and white specks to appear in the image. These artefacts have no correlation to the surrounding pixels, and so dominate the Laplacian metric. Due to JPEG encoding splitting the image into blocks, the JPEG metric is not similarly disrupted. Only the 8$\times$8 blocks containing the failed pixels were affected, making the JPEG sharpness metric more resilient to localised capture failures.

Using down-sampled resolution images, the Laplacian metric shows a clear peak centred in the range of focused images. The peak is noisy, but narrow and clearly higher than the background sharpness of unfocused images. By contrast, the Laplacian metric fails using full resolution images. There are numerous, single image peaks outside the focused range. These false peaks are only seen in the full resolution images, where the increased pixel count leads to a higher probability of a small area of dead pixels, which dominate the Laplacian score. The Laplacian convolution used employs a 3$\times$3 matrix, meaning only adjacent pixels are compared.  Using a 40$\times$ objective, the OpenFlexure Microscope has a field of view of $\sim$ 320 $\times$ \SI{240}{\micro\meter}. Capturing a full resolution image on the Pi Camera, the digital resolution of the image is 3280 $\times$ 2464 pixels. This corresponds to a physical pixel size $\sim$ \SI{0.1}{\micro\meter}, below the diffraction limit of high NA imaging. This oversampling combined with a 3$\times$3 Laplacian means that a significant amount of the contrast detected between neighbouring pixels is due to noise rather than focus. This issue becomes more severe with higher magnification imaging, where the narrower field of view ($\sim$ 140 $\times$ \SI{105}{\micro\meter}) is significantly oversampled by full resolution images.

The JPEG metric performed better; both the down-sampled and full resolution images displayed a clear, narrow peak centred in the focused range, with no false maxima. The JPEG size increases gradually and symmetrically either side of the peak, and remains consistent at a background size when the sample is fully out-of-focus. Using a 40$\times$ objective, the 8$\times$8 blocks in a full resolution image represent a $\sim$\SI{0.8}{\micro\meter} $\times$ \SI{0.8}{\micro\meter} area, which is more inline with the resolution of the microscope.

\begin{figure}
   	\centering
	{\includegraphics[width=0.49\textwidth]{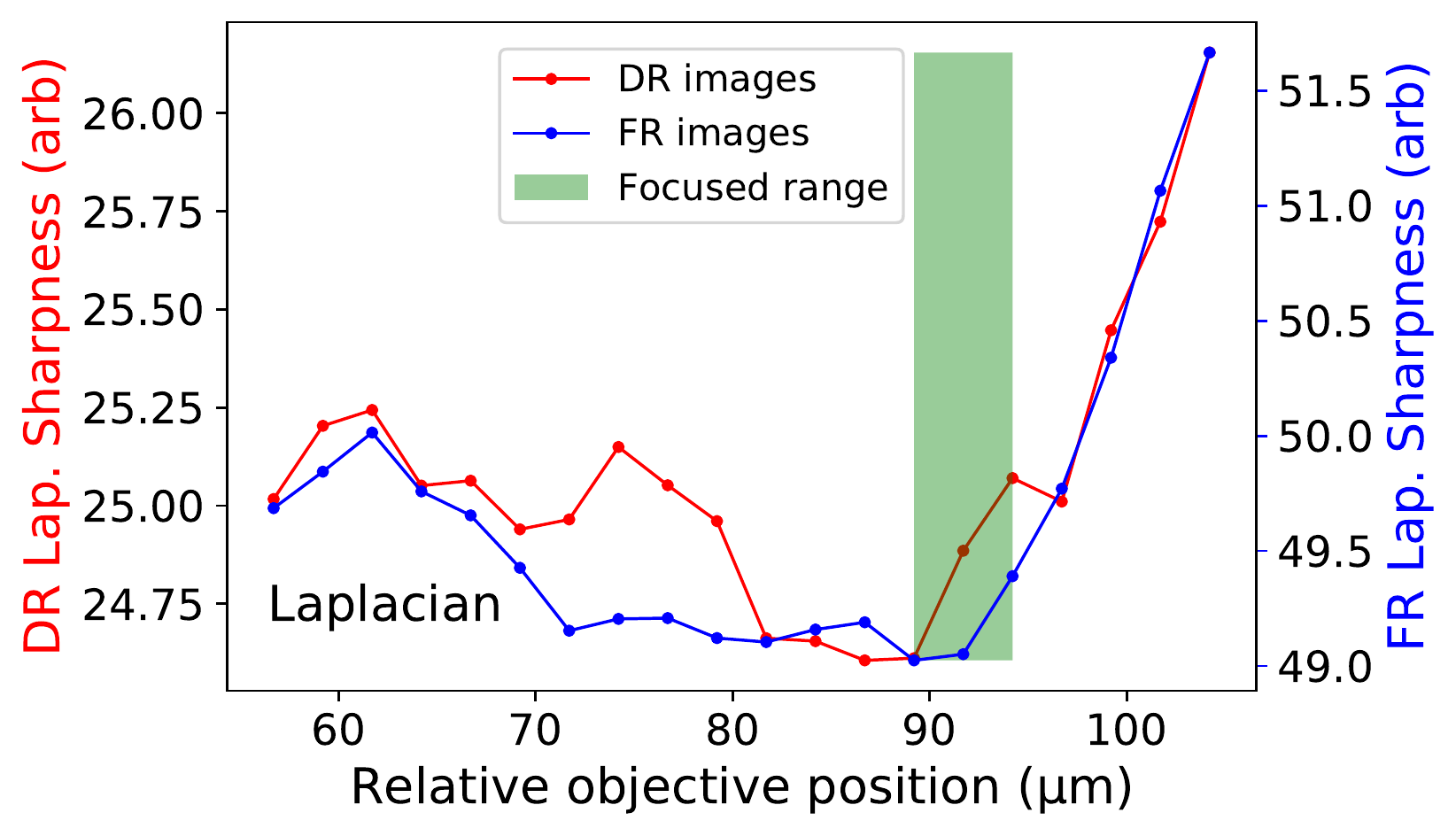}}
	{\includegraphics[width=0.49\textwidth]{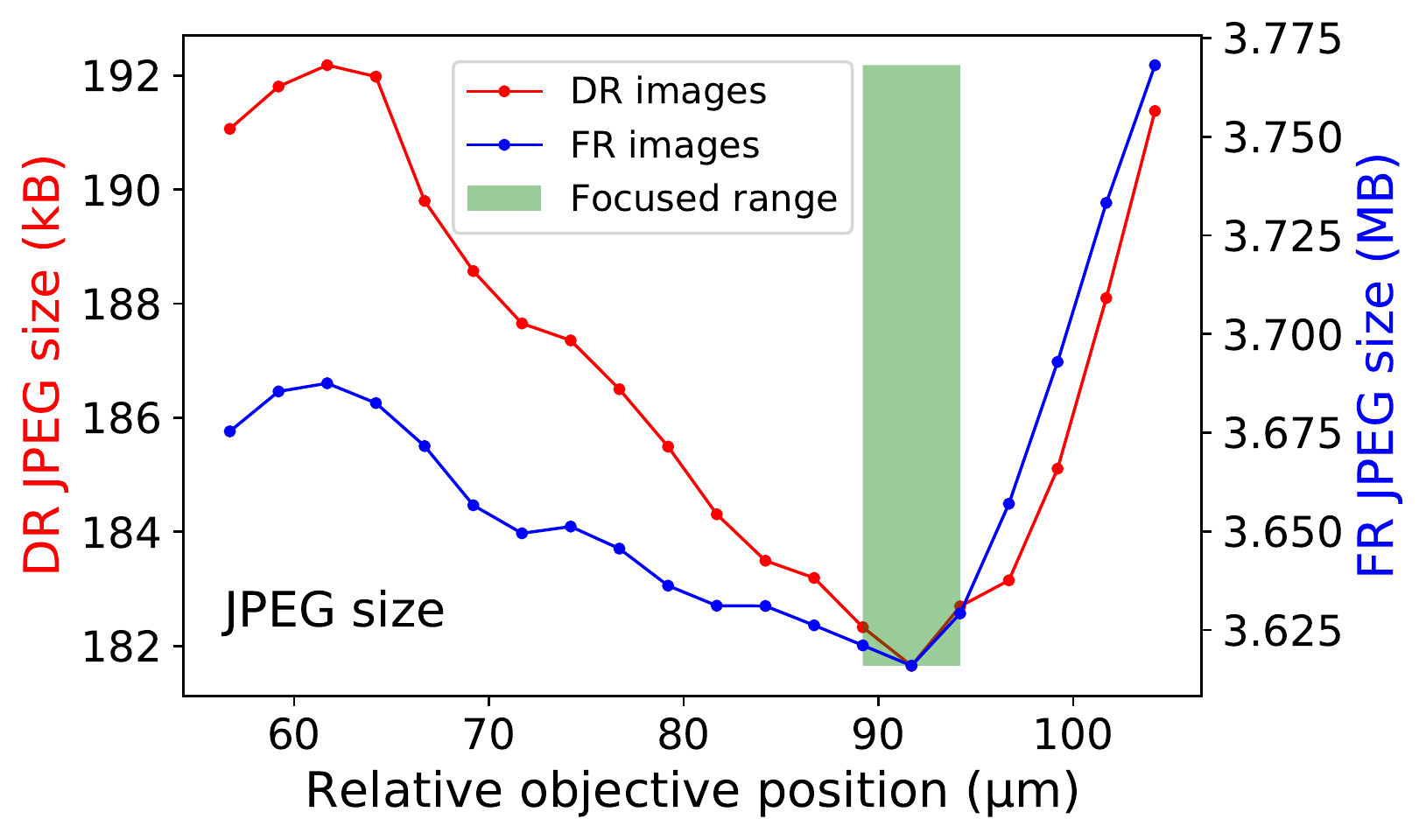}}
	\caption{A plot of JPEG size against objective height for a stack of images centred on the target shown in Figure \protect\ref{figure:USAF}. Both down-sampled resolution (DR) and full resolution (FR) image have their minimum in the manually judged focused range. Each step is \SI[separate-uncertainty = true,multi-part-units=single]{50(2)}{\nano\meter}.}
	\label{figure:inverted}
\end{figure}

Although the JPEG metric was found to be faster and more reliable on the samples tested, the division of the image into smaller blocks causes a potential failure state. On sufficiently sparsely-populated samples, a focused image may have the majority of blocks without features and showing only the background. Defocusing the image may cause information to ``overflow'' from populated blocks into these otherwise empty blocks. At this point, the defocusing begins to introduce more information to the image, increasing the file size. Examples of focused and defocused images of a sparse sample are shown in Figure \ref{figure:USAF}. Figure \ref{figure:inverted} shows a graph of JPEG size against height for this sample, where the minimum JPEG size is within the focused range. This would lead the JPEG metric to cause an autofocus to fail on this sample, instead positioning the sample far from focus.

This failure mode only affects sparse samples, such as individual latex beads or a single hard edge. As these failures are uncommon and understood, the JPEG metric is included in OpenFlexure software, with documentation recommending its use be avoided on such samples \cite{OFMserver,OFMDocs}. As the Laplacian metric averages pixel contrast over the full image, it is more reliable for less populated samples.

\section{Hardware positioning}
\label{section:moving}

When travelling in a single direction, the OpenFlexure Microscope translation stage is highly precise. The printed mechanism turns a single half-step of the stepper motor into an average movement in $xy$ of {88 $\pm$ 6 nm}. To provide the accuracy in focusing required for high magnification imaging, the $z$ axis mechanism has a greater resolution of \SI[separate-uncertainty = true,multi-part-units=single]{50(2)}{\nano\meter} per half-step. Error in the position is introduced, however, when the motors change direction. Mechanical backlash in the motor's internal gearing and the printed gears connecting it to the leadscrew introduce a slight lag between the motors and stage changing direction. This backlash reduces the certainty in position, but can be estimated by measuring the average distance moved by the motors before sample movement is observed. 15 trials in $xy$ indicated the OpenFlexure Microscope has an average backlash of 137 $\pm$ 19 steps, corresponding to an uncertainty in $xy$ of 12.1 $\pm$ \SI{1.9}{\micro\meter} and 6.9 $\pm$ \SI{1.0}{\micro\meter} in $z$. Measurements of the $xy$ movements and field of view were performed using OpenFlexure's calibration code \cite{Micat, OFMserver}. Measurement of the $z$ step size was based on a procedure used previously to measure OpenFlexure precision \cite{Grant2020}.

Although this effect is minor compared to the field of view in the plane of the sample, \SI{6.9}{\micro\meter} in $z$ is far greater than the depth of field of the 100$\times$ objective used in blood imaging. This means that the microscope cannot reliably measure the sharpness in a single movement, then return directly to the measured focal point. The change of direction would introduce uncertainty greater than the full range of the z-stack. This means not only would the central image not necessarily be at the focal point, but the z-stack may miss the focal point entirely. Including additional movements to always approach the focal plane from the same direction mitigates the backlash for imaging on systems with objectives between 4$\times$ and 40$\times$, but the lower depth of field of 60$\times$ and 100$\times$ objectives is close enough to the backlash uncertainty of $\pm$ \SI{1.0}{\micro\meter} to require further correction.

\begin{figure*}
   	\centering
	\includegraphics[width=0.85\textwidth]{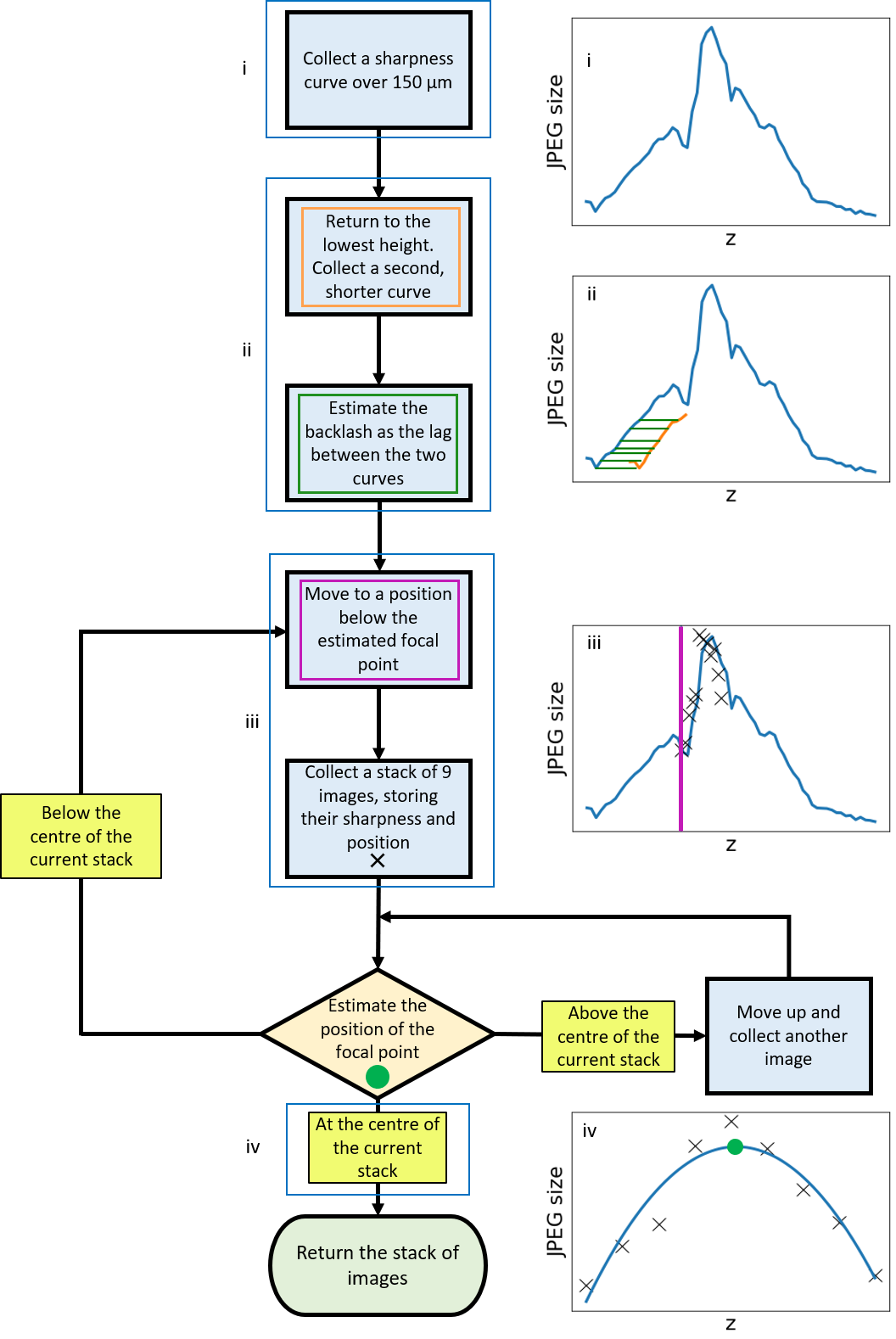}
	\caption{A flowchart of the movements, measurements and captures performed in the smart z-stack. Steps to estimate the optimal distances and intelligently abort the procedure may be added.}
	\label{figure:flowchart}
\end{figure*}

A ``smart stack'' procedure \cite{Stack2021} was devised and implemented in Python, using the OpenFlexure Python library \cite{PyClient2021}. The new procedure performs movements and measurements in a more reliable order. It utilises the speed and reliability of the JPEG sharpness metric, as the blood samples used in malaria diagnosis are densely populated enough to reliably work with this method. The basis of the smart stack is to use the sharpness measurements to estimate backlash and verify that the 9-image stack has its most focused image towards the centre. After estimating the distance between the objective and the focal plane of the sample, the microscope will only move in a single direction in $z$, eliminating the effect of backlash. An overview of the procedure and example measurements are shown as a flowchart in Figure \ref{figure:flowchart}.

The first step is to perform an initial sweep around the focal point, moving the objective from below the focal point to above to build up a sharpness against $z$ curve. Moving from one $xy$ grid position to the next, the microscope is only expected to move a maximum of few microns out of focus, and so this sweep is centred at the height of the previous stack. The FWHM of the sharpness peak for a blood sample is $\sim$ \SI{25}{\micro\meter}, and over $\sim$ \SI{75}{\micro\meter} the peak lowers to a background sharpness on both sides. To ensure these details are captured, the calibration curve is collected over a range of \SI{150}{\micro\meter}. The curve is then tested to ensure there is a sufficiently tall, narrow peak in the range. If this peak is absent or appears misshapen, a warning is recorded, as this may indicate an issue with the movements or positioning of the sample.

\begin{figure*}
   	\centering
	\includegraphics[width=0.95\textwidth]{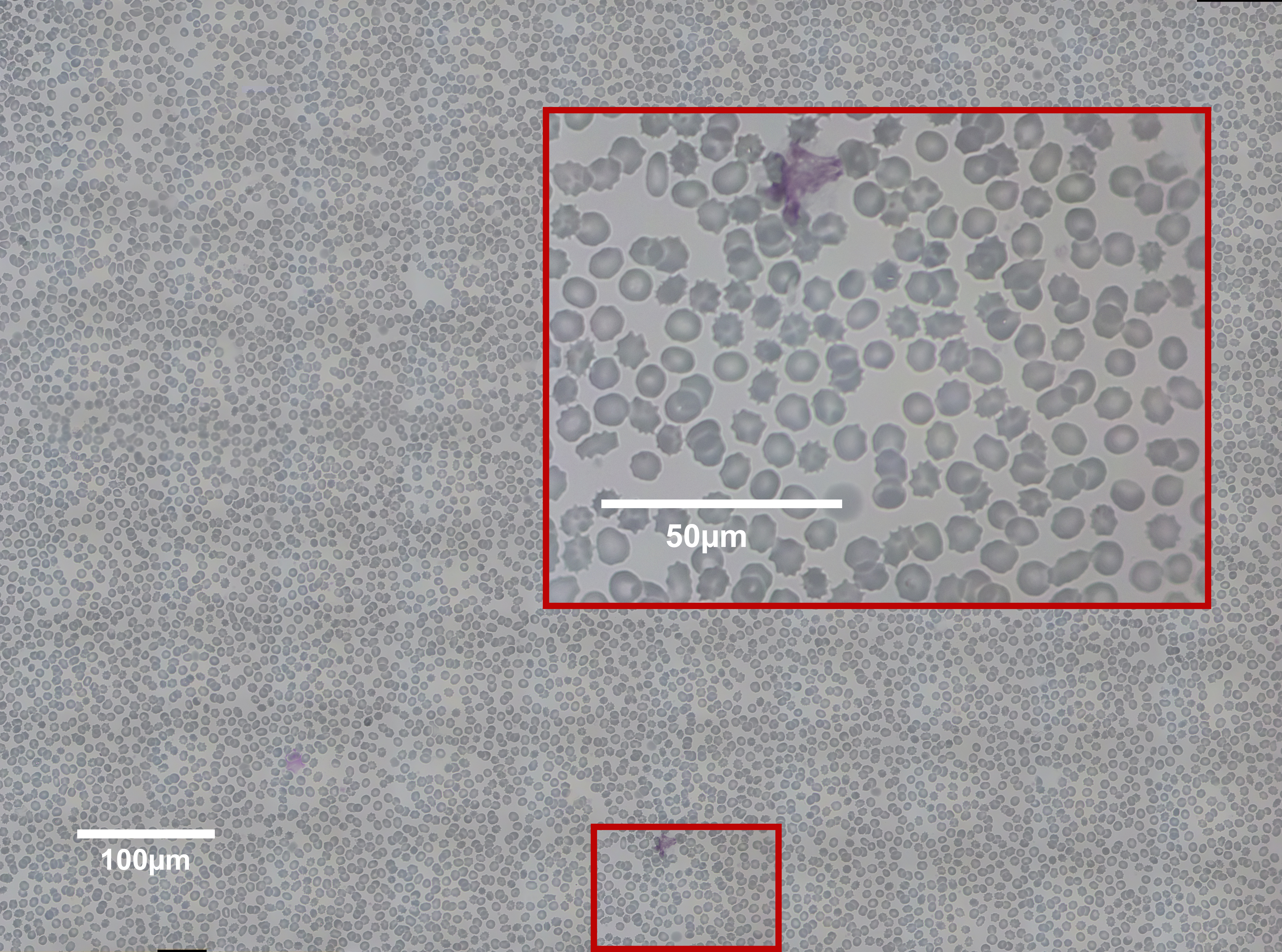}
	\caption{Tiled scan image of a Giemsa-stained thin blood smear, obtained with a 100$\times$, 1.25 NA oil immersion objective and focused using the smart stack algorithm. The composite image was built from a 10 $\times$ 10 grid of full resolution captures with an approximately 40\% overlap, using the Fiji stitching plugin \cite{Preibisch2009}. Inset: a single capture used to assemble the tiled image.}
	\label{figure:tiled}
\end{figure*}

As the sharpness curve varies depending on the direction of travel, the objective is then returned to below the focal point. A second, shorter curve is collected, over a range sufficient to show the trend in sharpness while remaining far from the estimated position of the stack. The lag between the two curves is estimated to predict any unexpected shift in positioning. The initial sweep and this lag is then used to make a second movement towards the start of the z-stack, below the focal point. To account for errors in movement, this movement is intentionally undershot, as the time penalty for overshooting and restarting is far greater than for undershooting.

The microscope then begins the z-stack, storing a series of images, their positions and their JPEG file sizes. The sizes are assessed against the peak of the initial sweep, to ensure the images captured have a similar sharpness to the estimated focal point. If so, a fourth-order Chebyshev polynomial fit is applied to the sharpnesses \cite{Harris2020a}. The sharpnesses and Chebyshev fit are then assessed using the following parameters:

\begin{itemize}
    \item the number and position of the turning points in the fit, as around the focal point there should be a single sharpness peak
    \item the gradient of the approach to and regression from the peak, to ensure the sharpness is peaking as expected
    \item the $z$ position of the peak within the stack, aiming to centre the stack on the focal point where possible
\end{itemize}

If the criteria indicate a well-focused z-stack centred on the focal point, the images are saved and the image closest to the peak of the fitted polynomial is marked as the focused image to use in tiling and diagnosis. The microscope then moves to the next $xy$ location. If the peak is estimated to be lower than the lowest point of the z-stack, the procedure is automatically restarted. If the peak is unclear or estimated to be above the z-stack central image, the microscope continues the z-stack, replacing the lowest image with one taken at the next $z$ position above the original stack. The shifted stack is then assessed as before. A break height far above the predicted peak is included as a failsafe, preventing the software from running indefinitely or causing the objective to strike the sample.

\section{Results}

The comparison of sharpness metrics showed that each are suitable for different applications. The JPEG size is quicker, less affected by noise and less computationally intensive, but unsuitable for sparse or plain samples. The Laplacian is more affected by noise and failures in captures, but works on the sparse or plain samples that cause the JPEG method to fail. Figure \ref{figure:sharpness} indicates that using the highest resolution of image available can cause the metric to be less reliable. The increased likelihood of a capture failure, and the image resolution exceeding the optical resolution of the microscope affects the 3$\times$3 Laplacian matrix to an unacceptable level. Lower resolution images are quicker to save and assess, require less storage space, and have a resolution more in line with the optical set up.

The failure caused by the 3$\times$3 Laplacian kernel testing for contrast below the diffraction limit may be mitigated by using a larger kernel which evaluates based on a larger area around each pixel. Limited computational resources, however, make this unviable for large area scans requiring many autofocus routines. Larger kernels are more computationally expensive, increasing the time required to focus.

Regardless of the choice of kernel size, the Laplacian sharpness metric requires each image to be captured, opened, converted to a Numpy array and converted to greyscale before the kernel is applied. These steps contribute a significant amount to the time taken performing the Laplacian sharpness metric. By contrast, the JPEG size metric requires only the file size from the GPU, and so can be performed in real time alongside continuous movement. This lack of a processing backlog means the required movements can be tested and fed into a closed loop feedback system without delaying results.

Although the faster JPEG metric is used to assess images, the use of the smart stack algorithm extends the duration of a 10$\times$10$\times$9 image blood scan from approximately 45 minutes to 80 $\pm$ 8 minutes. Some delay is expected, as the smart stack algorithm follows the original stack procedure, but allows the possibility of extra images being taken, or a stack being restarted if it appears to fail. The improved reliability of scans compensates for this delay however, by decreasing wasted time retaking images or scans when the z-stack fails. An example of a 10 $\times$ 10 image scan of a blood sample using the smart stack procedure is shown in Figure \ref{figure:tiled}. All images were captured automatically using the OpenFlexure Microscope, using a generic plan corrected achromatic 100$\times$ oil immersion objective with an NA of 1.2.

\begin{figure}[t]
   	\centering
   	\includegraphics[width=0.35\textwidth]{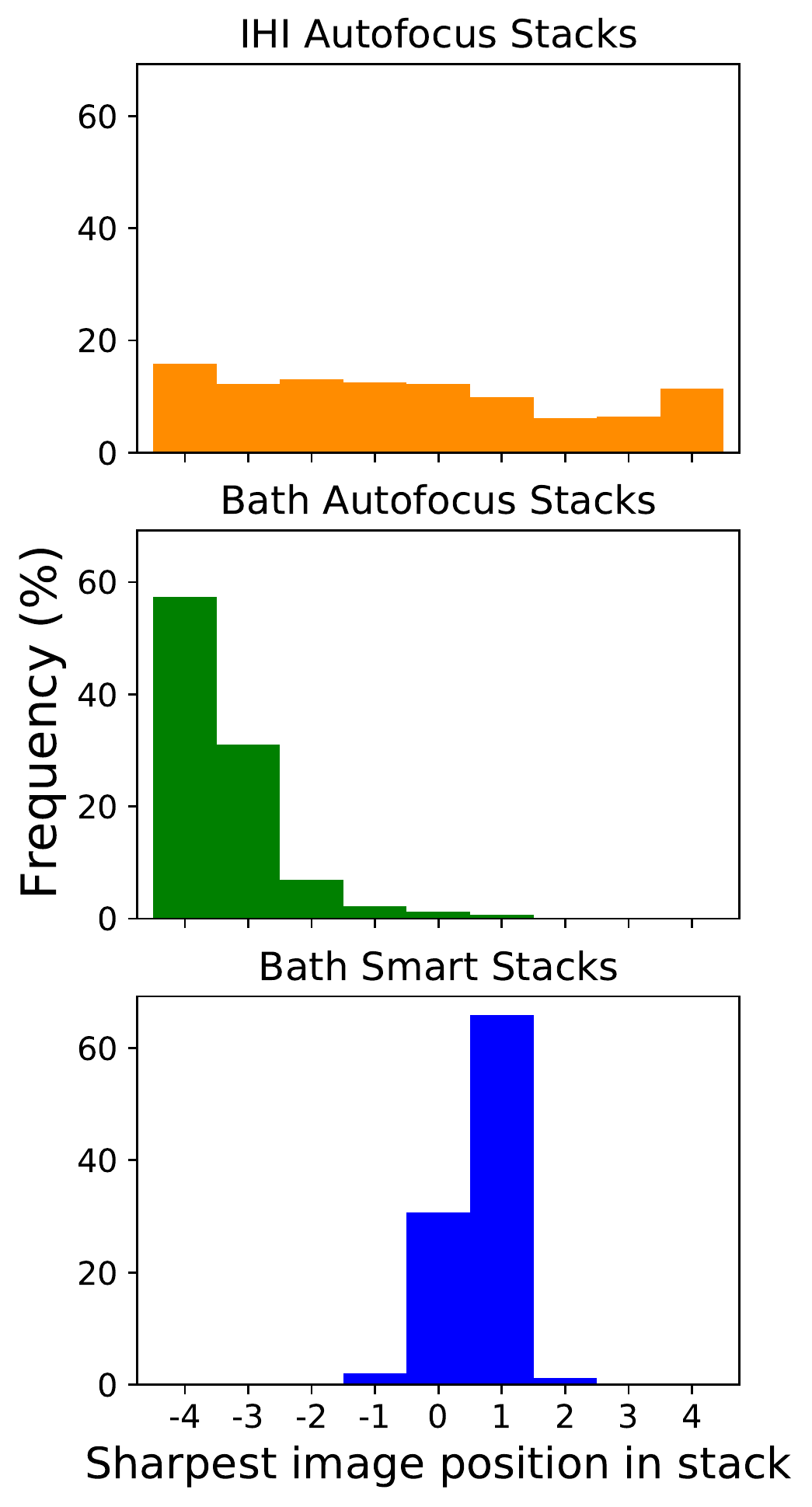}
	\caption{Comparison of the position of the sharpest image in a 9 image z-stack with d$z = $ \SI{0.5}{\micro\meter}. Stacks tested were from historical autofocus-only stacks from Tanzanian clinics (IHI), autofocus-only stacks collected in Bath, and smart stacks collected in Bath. Number of stacks tested = 59,427 IHI, 4,100 Bath autofocus and 1,127 smart stacks.}
	\label{figure:comparison}
\end{figure}

The improved reliability is shown in Figure \ref{figure:comparison}, by testing the position of the sharpest image in a stack of 9 images with d$z =$\SI{0.5}{\micro\meter}. If the autofocus and stack was successful, the sharpest image would be the centre of the stack. The historical data using JPEG autofocus but no smart stack was gathered from blood samples on OpenFlexure Microscopes running in Tanzanian health clinics (number of stacks tested $N = 59,427$) and in Bath ($N = 4,100$). The stacks from Tanzania have a similar likelihood of having the sharpest image at any position in the stack, with a skew towards the lower images. More than $57\%$ of the stacks taken in Bath had their sharpest image at the lowest image in the stack, a position that would be rejected by the smart stack procedure. Manual examination of these scans showed that in some cases the stack was entirely missing the focal plane, taking 9 unfocused and unusable images.

The smart stack data was collected on an OpenFlexure Microscope in Bath using multiple blood samples ($N = 1,127$). As with the historical data, images were collected using a 100$\times$ objective and a stack of 9 images with d$z =$ \SI{0.5}{\micro\meter}. Over 95\% of stacks had the sharpest image in the central three images (range = \SI{1}{\micro\meter}), showing the stack had centred around a peak sharpness. This indicates a significant improvement in identifying and centring on the focal plane of the sample.

\section{Implementation on the OpenFlexure Microscope}

The findings of this paper, including sharpness metric suitability, effects of backlash and effects of performing a smart stack, are included in the OpenFlexure documentation \cite{OFMDocs}. Users will be informed of when it is more suitable to use the JPEG size or Laplacian metric, along with troubleshooting for unexpected failures. This allows the end user to make their own informed decision, based on their sample, setup and requirements of reliability against speed. Both metrics are available in the OpenFlexure software, and smart stack will be included as the blood scan method in a future release.

\section{Conclusion}

The advantages of Laplacian and JPEG size sharpness metrics have been explored in the context of the sample and available computing resources. Trials have indicated that these metrics can be made quicker and more reliable by assessing images at a lower resolution. Based on the nature of blood samples and the level of focusing precision needed for high-power imaging, the JPEG metric was selected as suitable for large-area blood scans required for malaria diagnosis.

Once a focal point has been estimated, mechanical backlash necessitates the inclusion of additional error-correction steps before movement, allowing the microscope to approach the focal point from a consistent direction. The open-loop stage control also necessitates the stacks of images to be tested as they are collected, to identify if they contain a peak resembling the sharpness expected around a focal point. The effect of including the calibration and tests have been demonstrated on the blood samples which are the focus of our development.

Improving the user experience of working with the OpenFlexure Microscope motivated the inclusion of both sharpness metrics and stack options in the OpenFlexure software. The advantages and drawbacks of each option are explained in the documentation, allowing users to make their own decisions based on their workflow and priorities. When performing autofocus in a clinical environment, the $>$95$\%$ confidence in a clear, well focused z-stack benefits the user, advancing the aim of supporting malaria diagnosis through automated smart microscopy.

\section{Acknowledgements}

We acknowledge financial support from EPSRC (EP/R013969/1, EP/R011443/1) and The Royal Society (URF\textbackslash R1\textbackslash 180153).

\noindent All data and analysis code created during this research is openly available from the University of Bath Research Data Archive at [doi to be confirmed during proofing].

\bibliography{library}

\end{document}